\documentstyle[eqsecnum,prd,aps,epsfig]{revtex}
\begin{document}
%
%
\def\ov{\over}
\def\l{\left}
\def\r{\right}
\def\be{\begin{equation}}
\def\ee{\end{equation}}
\draft
\title{Neutron star transition to strong-scalar-field state in
tensor-scalar gravity}
\author{J\'er\^ome Novak}
\address{D\'epartement d'Astrophysique Relativiste et de Cosmologie \\
  UPR 176 du C.N.R.S., Observatoire de Paris, \\
  F-92195 Meudon Cedex, France}
\date{\today}
\maketitle

\begin{abstract}
Spherical neutron star models are studied within tensor-scalar
theories of gravity. Particularly, it is shown that, under some
conditions on the second derivative of the coupling
function and on star's
mass, for a given star there exist two strong-scalar-field solutions as
well as the usual weak-field one. This last solution happens to be
unstable and a star, becoming massive enough to allow for all three
solutions, evolves to reach one of the strong field
configurations. This transition is dynamically computed 
and it appears that the star radiates away the difference in energy
between both states (a few $10^{-3} M_\odot c^2$) as gravitational
radiation. Since part of the
energy ($\sim 10^{-5} M_\odot c^2$) is injected into the star as kinetic
energy, the velocity of star's surface can reach up to $10^{-2}
c$. The waveform of this monopolar radiation is shown as well
as the oscillations undergone by the star. These oscillations are also
studied within the slowly-rotating 
approximation, in order to estimate an order of magnitude of the
resulting quadrupolar radiation. 

\end{abstract}


\section{Introduction}
\label{s:intro}

Tensor-scalar theories have been studied in many works (see
e.g.\cite{DEF92}) as natural generalizations of Einstein's general
relativity, representing the low-energy limit of superstring theories
(\cite{CFM85} and \cite{DP94}). They describe gravity by the usual
spin-2 field ($g_{\mu\nu}$) combined with one or several spin-0 fields
($\varphi$). In all this paper, only one scalar field will be
considered, with no self-coupling (potential). In that case, the
most general tensor-scalar theory is given by the action:
\be
S=(16\pi G_*)^{-1}\int d^4x\,\sqrt{-g_*}(R_*-2g_*^{\mu \nu}\partial_\mu 
\varphi \partial_\nu \varphi) + S_m[\Psi_m,a^2(\varphi)g^*_{\mu \nu}],
\ee
where all quantities with asterisks are related to the
``Einstein metric'' $g^*_{\mu \nu}$: $G_*$ is the bare gravitational
 coupling constant,
 $R_*=g_*^{\mu \nu}R^*_{\mu \nu}$ the curvature scalar for this metric and
 $g_* = \det(g^*_{\mu\nu})$. The term $S_m$ denotes the action of matter,
represented by the fields $\Psi_m$, which is coupled to the ``Jordan-Fierz''
metric $\tilde{g}_{\mu \nu}= a^2(\varphi)g^*_{\mu\nu}$; all quantities with
a tilde are related to this metric. That means that all non-gravitational
experiments measure this metric, although the field equations of the
theory are better formulated in the Einstein one. By varying $S$, one
obtains 
\begin{eqnarray} 
 && R^*_{\mu \nu} - {1\ov 2}g^*_{\mu \nu}R^*  = 2\partial_\mu \varphi
\partial_\nu \varphi - g^*_{\mu \nu}g_*^{\rho \sigma} \partial_\rho
\varphi \partial_\sigma \varphi + {8\pi G_*\ov c^4} T^*_{\mu \nu}, 
\label{e:tscal}\\
 && \Box_{g_*} \varphi = -4\pi G_* \alpha(\varphi)T_* , \label{e:ondsc}
\end{eqnarray}
 where the Einstein-frame stress-energy tensor
$T_*^{\mu\nu}=2(-g^*)^{-1/2}\, \delta S_m/\delta g^*_{\mu\nu}$ is related
to the physical one, $\tilde{T}^{\mu \nu} = 2(-\tilde{g})^{-1/2}\,
\delta S_m / \delta  \tilde{g}_{\mu \nu}$, by
\be
T^\mu_{*\nu} = a^4(\varphi) \tilde{T}_\nu^\mu .
\ee
The logarithmic derivative
of the coupling function $a(\varphi)$ is $\alpha(\varphi)$, present in
eq.~(\ref{e:ondsc}). It represents the
field-dependent coupling strength between matter and the scalar
field. Hereafter, it is assumed that this coupling strength function
contains no large dimensionless parameter, and hence the class of
coupling functions is well represented by an affine one, depending on
two parameters (see \cite{DEF96}). If $\varphi_0$ denotes the
asymptotic value of the scalar field, 
 $\alpha_0=\alpha(\varphi_0)$ and $\beta_0=\partial \alpha(\varphi_0) /
\partial \varphi_0$ can be chosen\footnote{$\varphi_0$ is then
redundant with $\alpha_0$ and does not represent a parameter for the
theory} to parametrize $\alpha(\varphi)$:
\be
\alpha(\varphi) = \alpha_0 + \beta_0 \times(\varphi - \varphi_0).
\label{e:cfunc}
\ee 
 Brans-Dicke theory is obtained for
 $\beta_0=0$, on the opposite, even if $\alpha_0=0$ (and $\beta_0 \neq 0$),
the scalar field can exhibit some non-perturbative effects in neutron
stars (strong gravity), when $\beta_0$ is lower than some critical
value (about -5, depending on the mass of the star). This phenomenon
has been described in \cite{DEF96} and \cite{DEF93} as ``spontaneous
scalarization'', by analogy with the spontaneous magnetization of
ferromagnets below the Curie temperature. The aim of this paper is to
determine under which conditions a neutron star shows such spontaneous
scalar field and how can the dynamical transition be. Resulting
gravitational radiation will also be discussed. The organization is as
follows: section~\ref{s:SScond} delimits the parameter space for
non-perturbative effects to appear and sets a maximal value for
$\beta_0$, section~\ref{s:ssdyn} shows computed evolution of unstable
stars, during which gravitational radiation is emitted
(section~\ref{s:gw}). Finally, section~\ref{s:conc} gives some
concluding remarks.

\section{Conditions for Spontaneous Scalarization}\label{s:SScond}

In all this section, spherical symmetry is assumed (this will not be
the case in sec.~\ref{ss:quad}), and the coupling function of the form
(\ref{e:cfunc}) 
will be considered throughout all the paper. It gives the following
expression for the function $a(\varphi)$:
\be
a(\varphi) = e^{\alpha_0 (\varphi - \varphi_0) + \beta_0/2(\varphi -
\varphi_0)^2} \label{e:afunc},
\ee
so that $a(\varphi_0) = 1$. As stated above, $\varphi_0$ enters the
theory only as a boundary condition for the scalar field, but not as a
parameter of the theory.

\subsection{Coordinate and Variable choices} \label{ss:coord}

Following \cite{NOV97}, space-time is decomposed on spacelike
hypersurfaces within the $3+1$ formalism, with the radial gauge and
polar slicing. The metric $g^*_{\mu \nu}$ takes the form
\be
ds^2 = -N^2(r,t)dt^2 + A^2(r,t)dr^2 + r^2(d\theta^2 +
 \sin^2\theta d\phi^2) \label{e:metric}
\ee
All coordinates are expressed in the Einstein-frame, and asterisks are
 omitted. However, ``physical'' quantities will often be written in the
Fierz metric and noted with a tilde. Neutron star matter is modeled as
a perfect fluid $$\tilde{T}_{\mu\nu} = (\tilde{e} + \tilde{p})
\tilde{u}_\mu \tilde{u}_\nu  + \tilde{p} \tilde{g}_{\mu\nu},$$ where
$\tilde{u}_\mu$ is the 4-velocity of the fluid, $\tilde{e}$ is the
total  energy density (including rest mass) in the fluid frame and
$\tilde{p}$ is the pressure. The description of the fluid is completed
by the equation of state, not depending on temperature (cold
matter)
$$
\tilde{e} = \tilde{e}(\tilde{n}_B),
$$
with $\tilde{n}_B$ being the baryonic density in the fluid
frame. This assumption consists in neglecting the strong and weak
nuclear interaction processes and assuming matter is at equilibrium
for these processes. This is, of course, not valid for
$\beta$-equilibrium, but as it has been shown by \cite{GH93}, the
effects of weak nuclear interactions on the hydrodynamics of a neutron
star collapse are negligible. Finally, the {\em coordinate velocity}
is defined as $$ 
V = {dr\ov dt} = {u^r \ov u^0},$$ and the Lorentz factor of the fluid as
$$\Gamma=\l(1-\l({A\ov N}V\r)^2\r)^{-1/2}.$$ The equations for the fields and
matter variables are those of \cite{NOV97} and are solved the same
way, by the same numerical code. There will be 4 equations of state
used in this work: 
{\it EOS1\/}, a polytrope ($\tilde{p} = K \tilde{n}_0
\tilde{m}_B\l({\tilde{n_B} \ov \tilde{n}_0}\r)^\gamma$) with 
 $\gamma=2.34$ and $K=0.0195$, $\tilde{m}_B=1.66 \times 10^{-27} \text{
kg}$ and $\tilde{n}_0 = 0.1 \text{ fm}^{-3}$; 
{\it EOS2\/}, the same as above, but with $\gamma=2$ and $K=0.1$;
{\it EOS3\/}, Pandharipande equation of state (realistic, rather soft
equation of state; see \cite{PAN71} and \cite{SBG94} for properties)
and 
{\it EOS4\/}, Bethe and Johnson equation of state (realistic, rather
stiff equation of state; see \cite{SBG94} and \cite{BJ74} for
properties).
These equations have chosen because they describe different
stiffnesses of matter and, on the other hand, their numerical behavior
allows for good accuracy in the results.

\subsection{Maximal $\beta_0$ parameter for Spontaneous Scalarization}
\label{ss:b0max}

When considering static solutions of
eqs.~(\ref{e:tscal})--(\ref{e:ondsc}) for $\alpha_0=0$, there may
exist two types of solutions:
 one with $g^*_{\mu \nu}$ being the same as in general
relativity, and $\varphi = \text{constant} = \varphi_0$, the scalar
charge of the star is thus null;
two solutions where $|\varphi-\varphi_0| \sim 1$ at star's center, 
they are images of each other by $\varphi\rightarrow-\varphi +
2\varphi_0$, and have the same global characteristics, in those cases,
the scalar charge is of the order of the star's mass.

\begin{figure}
\centerline{ \epsfig{figure=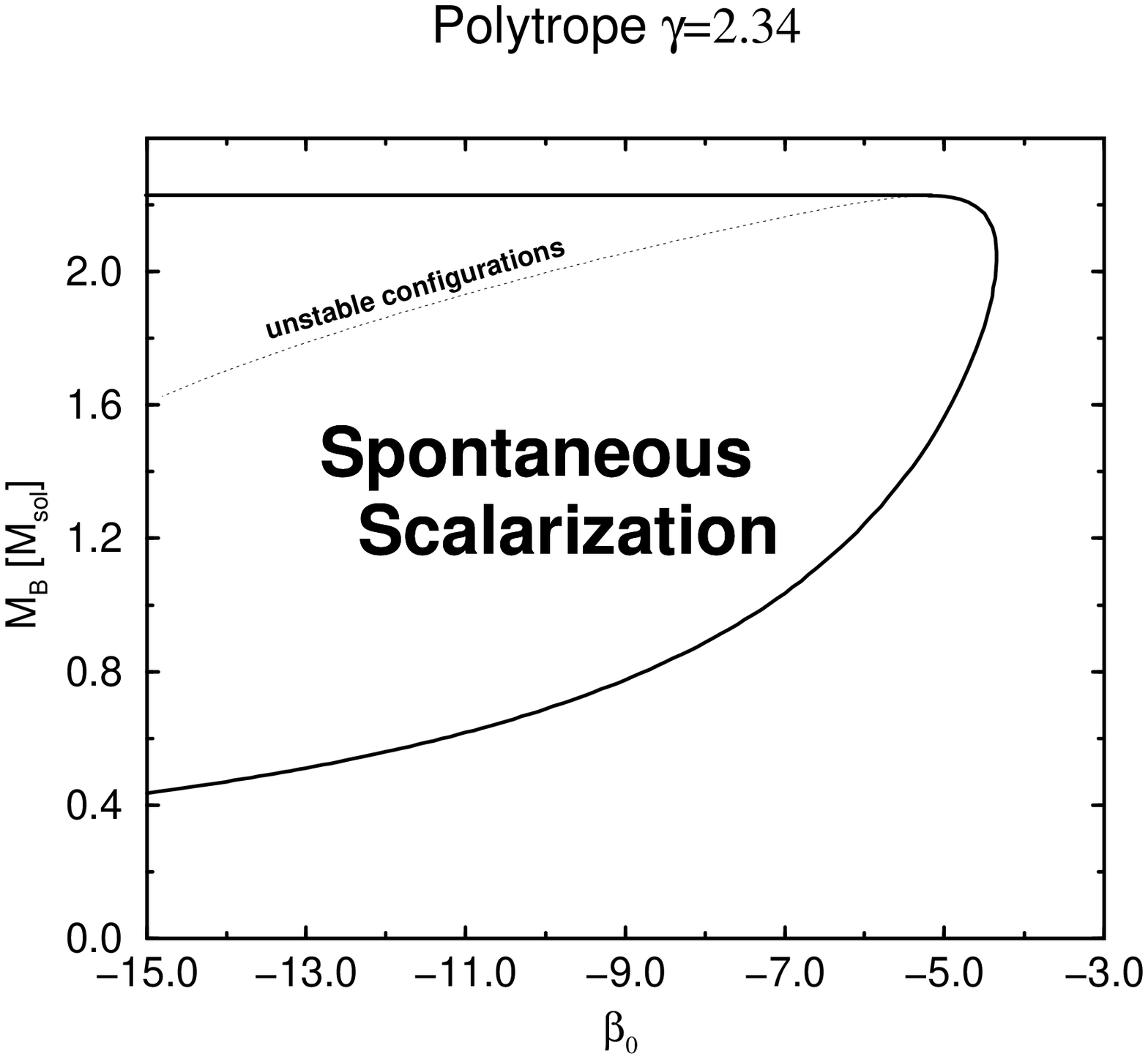,height=11cm,width=16cm,angle=0} }
\caption[]{\label{f:sspol1}   
Zone of spontaneous scalarization of general-relativistic neutron star
solutions in the baryonic mass ($M_B$) -- $\beta_0$ plane, for EOS1. The zone
lies inside solid lines, the horizontal line at $2.23 M_\odot$
represents the maximal mass for neutron stars in general
relativity. Since the curve is parametrized by stars' central density
($\tilde{n}_B$), it has been continued for unstable configurations
(thin dotted line) for which $dM_B/d\tilde{n}_B<0$. See
Sec.~\ref{ss:b0max} for more explanations}
\end{figure}

The first type of solution (which will be noted weak-field solution) always
exists, whereas the second one (called strong-field solution) requires
that the
amount of baryons in the star (its baryonic mass $M_B$) be larger than
some critical value, depending on the $\beta_0$ parameter. This result
has been obtained by Damour and Esposito-Far\`ese (\cite{DEF96},
Fig.~2 and Tab.~1) and Harada (\cite{HAR97}, Tab.1). Using the same
equation of state as these two works (EOS1), static weak-field
solutions were computed for 
increasing baryonic masses and checked against spontaneous
scalarization. Since in the $\alpha_0=0$ case the transition has an
infinite slope (Cf. Fig.~1 of \cite{DEF96}), it is possible to draw
the curve $M_B^{crit}$ as a function of $\beta_0$; the results are
displayed in Fig.~\ref{f:sspol1}. The curve is parametrized by stars'
central density, until it reaches its maximal value, corresponding to
maximal  mass  ($2.23 M_\odot$ for general relativity). If this
density is further increased, neutron stars configurations become
hydrodynamically unstable, they may however show spontaneous
scalarization effects if their mass is above the dotted curve.
For hydrodynamically stable neutron stars (being below the $M_B=2.23
M_\odot$ line), strong-field  solutions can develop {\em inside} the solid
curve. This is in accordance with previous works (\cite{DEF96} and
\cite{HAR97}) and shows that there exists a maximal $\beta_0$
parameter for spontaneous scalarization to occur ($\beta_0 \simeq
-4.34$). For $-5.2 < \beta_0 \leq -4.34$, Fig.~\ref{f:sspol1} shows
that a star, which is close to its maximal mass, does not exhibit
spontaneous scalarization effects, whereas for the same $\beta_0$ less
massive stars do. An explanation to this could be that, for such
stars $\tilde{e}-3\tilde{p}$ at their center becomes small or
negative, therefore the simplified model of static equations of
\cite{DEF93} shows no more ``zero modes''. However, this cannot be the
only reason, the $\beta_0$ parameter should also intervene.

\begin{figure}
\centerline{ \epsfig{figure=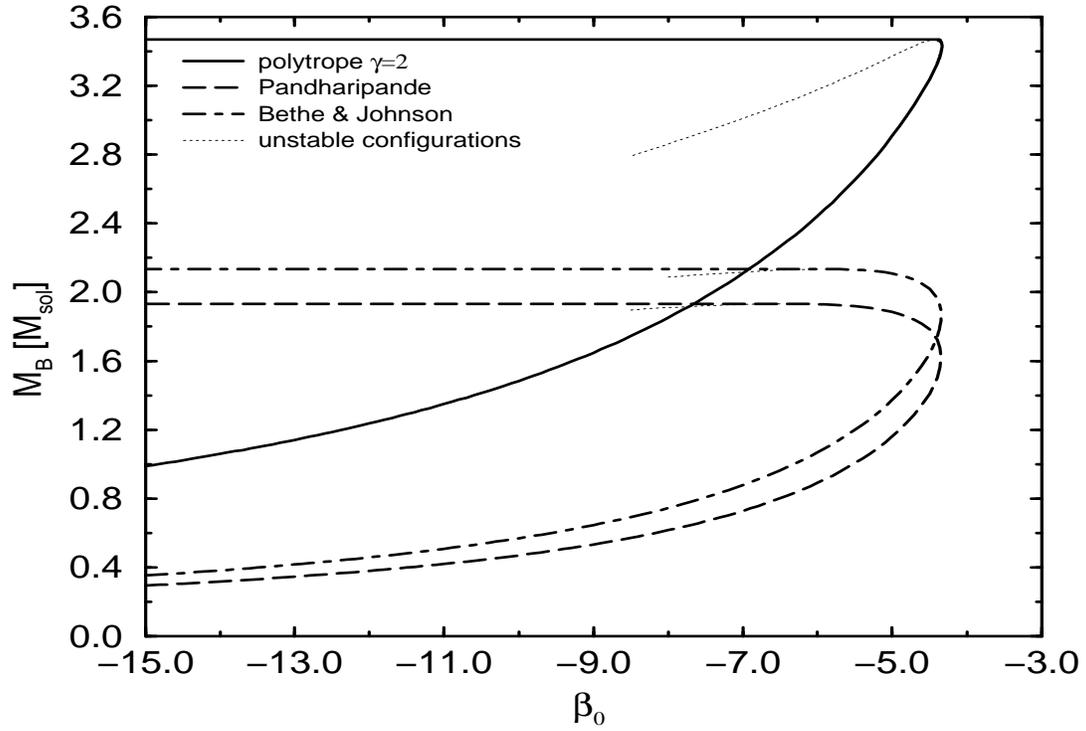,height=11cm,width=16cm,angle=0} }
\caption[]{\label{f:sseos}   
Same as Fig.~\ref{f:sspol1}, but for EOS2, EOS3 and EOS4 (see
Sec.~\ref{ss:coord}).}
\end{figure}

This study has been extended to other equations of state (EOS2 to 4)
and the results are shown in Fig.~\ref{f:sseos}. The main
result is that the maximal value of $\beta_0$ for which spontaneous
scalarization still occurs is independent from the equation of state
used. Its value reads 
\be
\beta_0^{\text{max}} = -4.34 \pm 0.01 \label{e:b0max}.
\ee
Since the equations of state used cover a very broad range of
stiffness and are, for two of them, results of realistic scenarios for
dense matter, this limit can be considered as a very strong one for
compact stars. It can easily be interpreted with the results of Harada
\cite{HAR97}: taking an incompressible fluid model, he has shown that
the $\beta_0$ parameter below which instabilities of the scalar field
develop, is little sensitive to the compactness ($R/M$) of the star
around $R/M=4$ (Fig.~5 of his work). Section~\ref{s:ssdyn} will show
that these dynamical instabilities are expression of the existence of
strong-field solutions.

\subsection{Spontaneous Scalarization when $\alpha_0 \not= 0$} 
\label{ss:a0n0}

What has been stated in previous section was a result of computations
in the case $\alpha_0=0$. Behavior of neutron stars is quite similar in
the case $\alpha_0 \not= 0$, if we consider this parameter constrained
by solar-system experiments (see \cite{DEF96} and \cite{GAMMA}) to: 
\be
\alpha_0^2<10^{-3} \label{e:a0lim}.
\ee
The most general conditions for spontaneous scalarization to appear
have been studied by \cite{HAR98}, using catastrophe theory. Namely,
for $\alpha_0$ verifying (\ref{e:a0lim}), there are still two kinds of
static solutions: one for which the 
scalar charge is of the order of $\alpha_0$, the second one
(containing two different solutions, not equivalent this time) for
which the scalar charge is of the order of unity. Some models of
neutron stars are described in Table~\ref{t:model} in order to compare
models with same number of baryons ($1.5 M_\odot$), with $\beta_0=-6$
for all of them so that spontaneous scalarization is likely to
occur. One sees that solutions with high scalar charge are
energetically more favorable than solutions with no or small charge;
beside this, spontaneous scalarization mainly affects matter
distribution inside the star, but very little global variables such as
the mass and the radius. One should note that in the case $\alpha_0\not= 0$,
when the mass of the neutron star increases, the transition to
spontaneous scalarization is smoothed and one has to
delimit the zone for spontaneous scalarization  to happen (as in
Fig.~\ref{f:sspol1}) by looking for the existence of three solutions
for a given amount of baryons. The study of the number of equilibrium
solutions and of their stability has been done, for EOS1, by Harada
\cite{HAR98}, using catastrophe theory. The same results have been
found here, using a static numerical code, and have been extended to
other equations of state, getting thus as a new result the maximal
value of $\beta_0$ (\ref{e:b0max}), compatible with spontaneous
scalarization in the $\alpha_0=0$ case.

\section{Dynamical transition to Spontaneous Scalarization state}
\label{s:ssdyn}

The conditions for spontaneous scalarization to appear are now well
defined and the question is now to know, for a fixed theory
(i.e. $\alpha_0$,  $\beta_0$ and the equation of state) and a
given amount of baryons, if the three equilibrium solutions are
stable and what happens to the unstable ones. The first part of this
question has already been answered by Harada \cite{HAR97} who did a
semi-analytical stability analysis of spherically symmetric neutron
stars in tensor-scalar theory; and by Harada \cite{HAR98} using the
catastrophe theory. He showed the development of unstable
modes for weak-field solutions which correspond to the possibility for
spontaneous scalarization. Here, the dynamical numerical code showed the
same results. This code is described in more details in \cite{NOV97},
the main point being the fact that, thanks to pseudo-spectral
techniques, it is precise enough to be sensitive to
instabilities, and to follow the evolution of the unstable star. The
stability results have been numerically checked,
and the code showed that the weak-scalar-field equilibrium solutions
were  unstable when strong-scalar field solutions existed
(e.g. solution 1 and 4 of Table~\ref{t:model}). 

\begin{figure}
\centerline{\epsfig{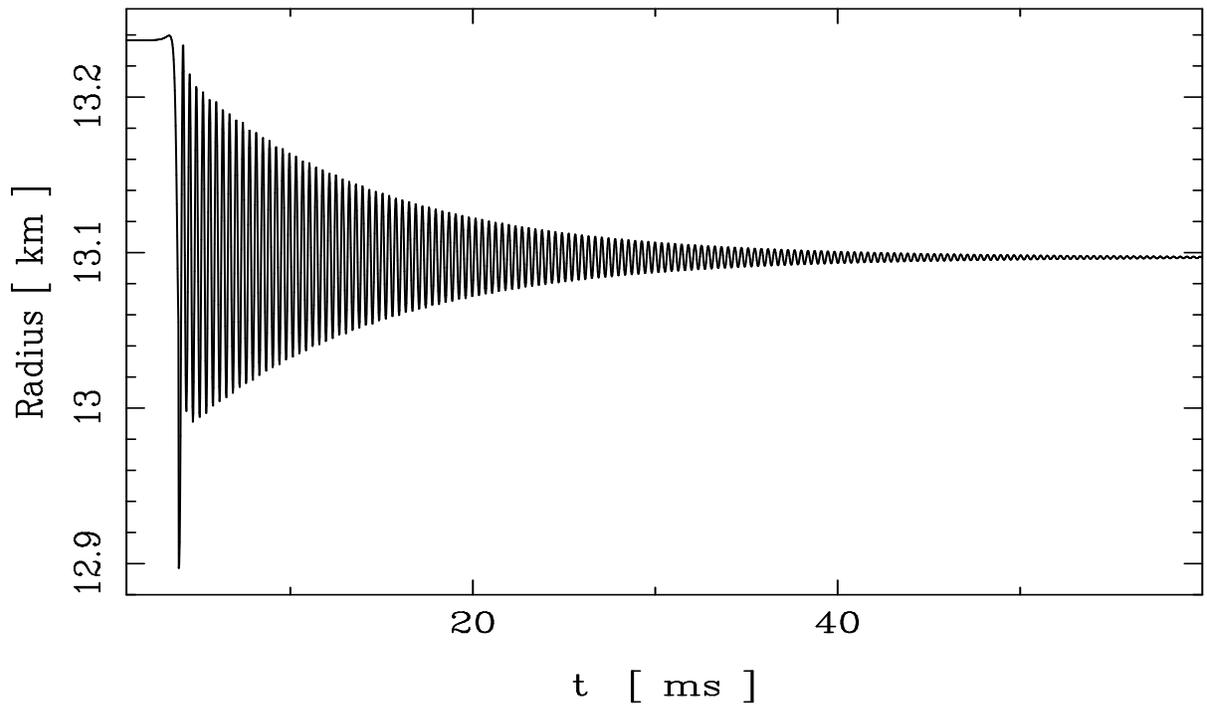}}
\caption[]{\label{f:varray}
Evolution of star's radius for unstable initial equilibrium solution number 4
(Cf. Table~\ref{t:model}) .}
\end{figure}

\begin{figure}
\centerline{\epsfig{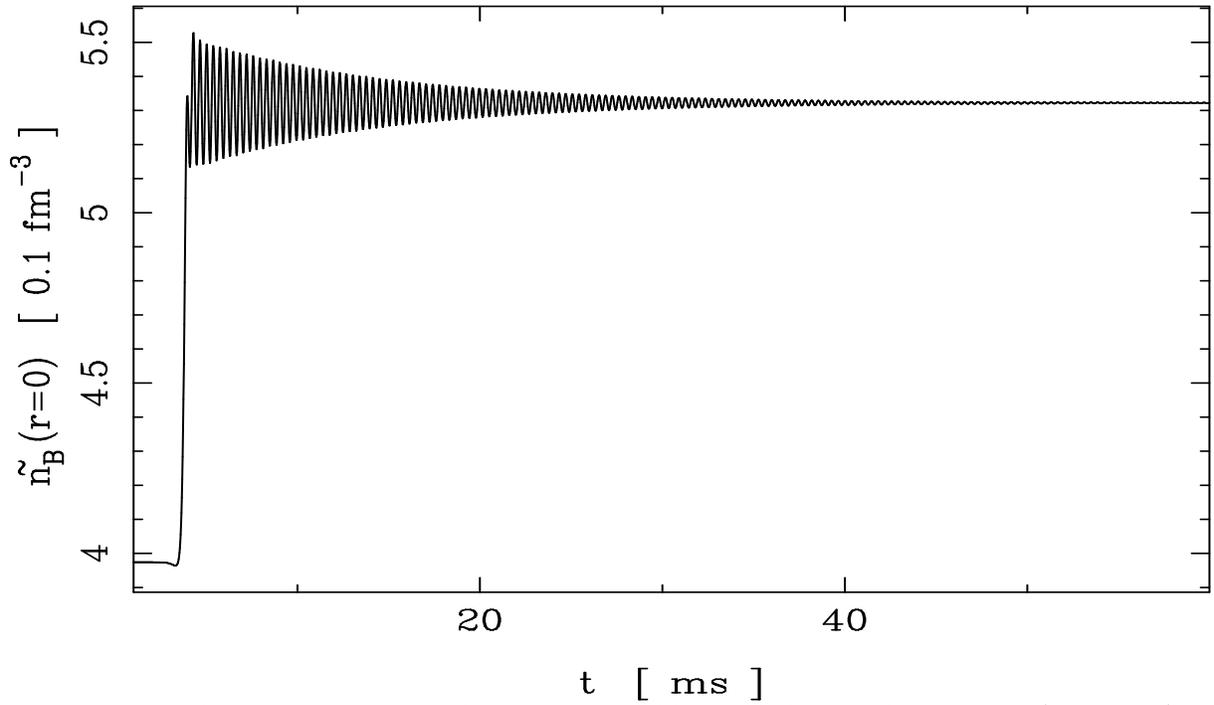}}
\caption[]{\label{f:varnb0}
Evolution of star's central density for unstable initial equilibrium solution
number 4 (Cf. Table~\ref{t:model}) .}
\end{figure}

\begin{figure}
\centerline{\epsfig{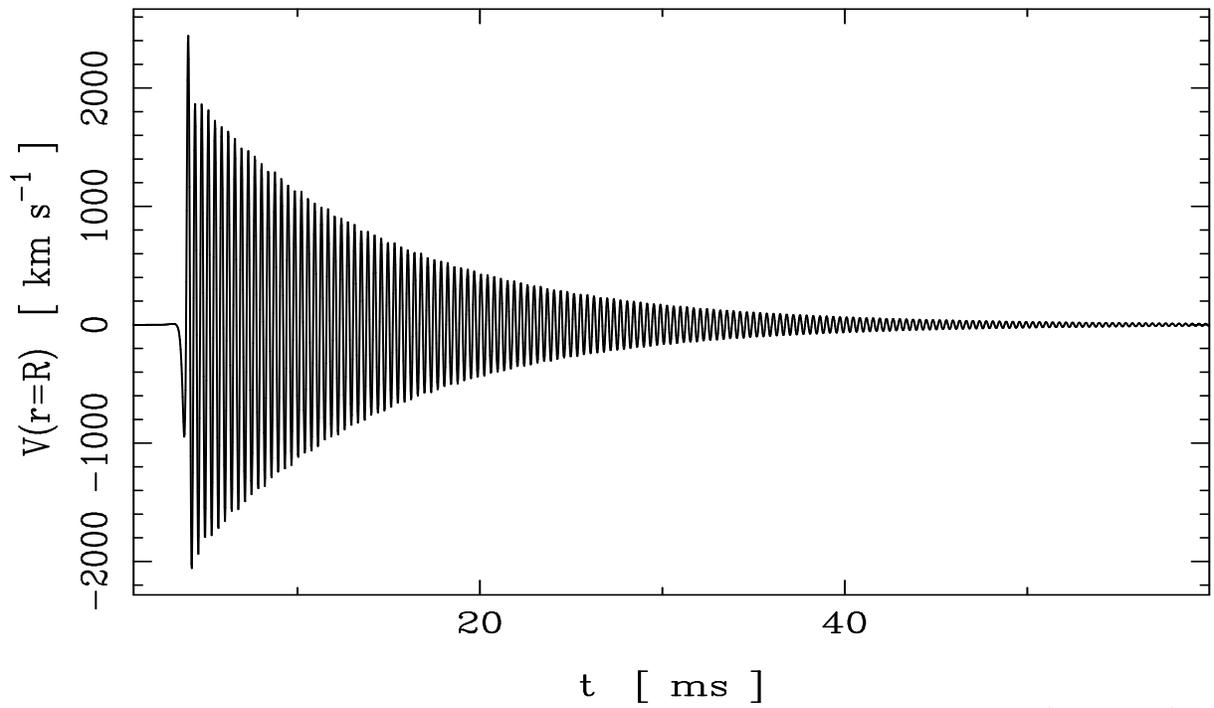}}
\caption[]{\label{f:varv}
Evolution of star's surface velocity for unstable initial equilibrium solution
number 4 (Cf. Table~\ref{t:model}) .}
\end{figure}

\begin{figure}
\centerline{\epsfig{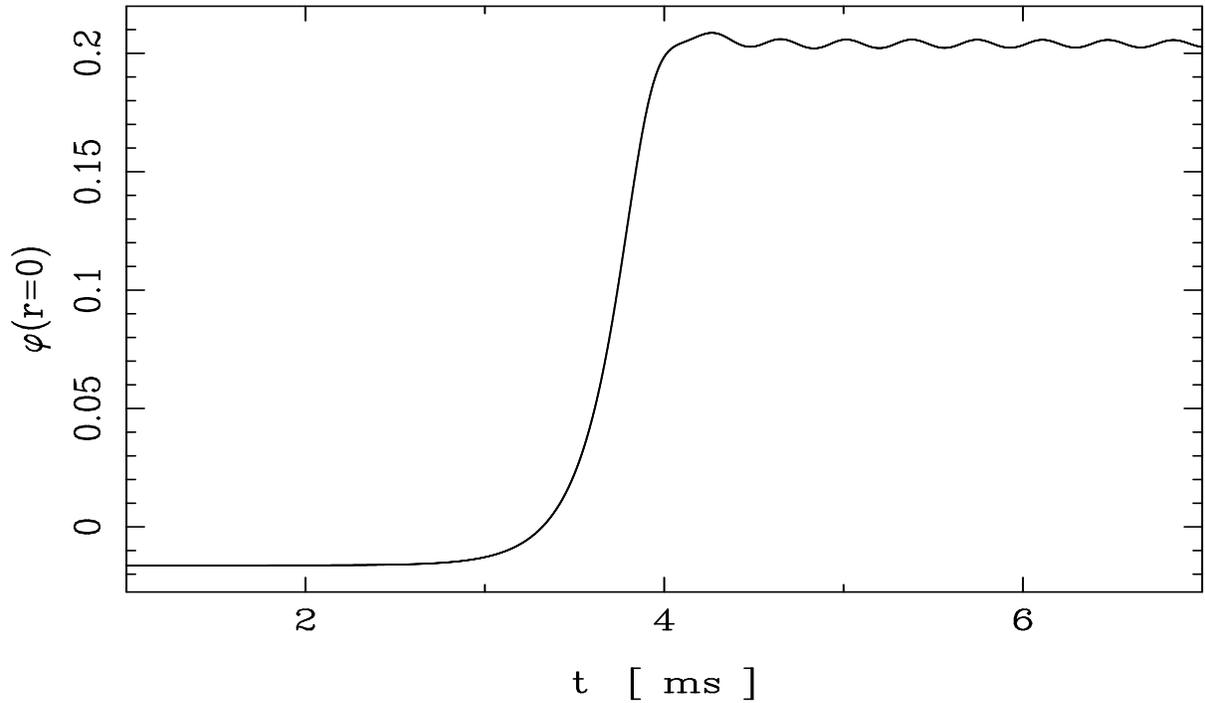}}
\caption[]{\label{f:varfi0}
Evolution of star's central scalar field value for unstable initial equilibrium
solution number 4 (Cf. Table~\ref{t:model}) .}
\end{figure}

This study can be linked with the physical scenario of an accreting
neutron star (e.g. in an X-ray binary system), which passes the limit
of the critical baryonic 
mass. When it is born, the star may be below this mass and thus either in
a general relativistic state (if $\alpha_0=0$) or in a
``weak-scalar-field'' state. When it passes the critical baryonic
mass, the star will be in an unstable configuration. This kind of solution
was then numerically followed and resulting evolution is shown on
Figs.~\ref{f:varray} to \ref{f:varfi0}, for radius, central density,
surface velocity and central scalar field, beginning with the unstable
solution number $4$ of Table~\ref{t:model}. One sees that the star
undergoes a strong variation of its matter distribution, caused by
the rapid raise of the scalar field, then behaves like a damped
oscillator radiating away its kinetic energy through monopolar
gravitational radiation (see Sec.~\ref{ss:mono}) and finally settles down
to the strong-field state (static solution 5 of
Table~\ref{t:model}). One can thus see, in the case $\alpha_0 \not=
0$,  the spontaneous scalarization
appearing dynamically. In the simulations, the star starting form the
weak-field 
configuration, would settle down either to the positive strong-field
state (solution 5) or to the negative one (solution 6). The
final dynamically-evolved fields correspond to those obtained from a
static code within $1\%$ of error (the code indicates $3\%$, see
\cite{NOV97}) and the baryonic mass is conserved up to $10^{-5}$.

Such simulations have been done for various masses and coupling
function parameters. Results were very similar: the weak-field state
 is always unstable when  the two
other strong-field states exist; these last solutions are both
stable. Looking at the energies, one sees that the weak-field solution
is a local maximum, whereas spontaneous scalarization states are local
minima (even if the state having a scalar charge of the opposite sign
to $\alpha_0$ is energetically less favorable than the other one, see
table~\ref{t:model}). The difference in energy is radiated away as
monopolar gravitational waves in two steps.
First, the largest part (about $99\%$) of the difference in energy
between both states is radiated very rapidly when the scalar field
grows to its new value;
then, the rest of the energy is put into the star as kinetic energy
which is dissipated slowly by the oscillations of the star and the
change they induce onto the scalar field.

\begin{figure}
\centerline{\epsfig{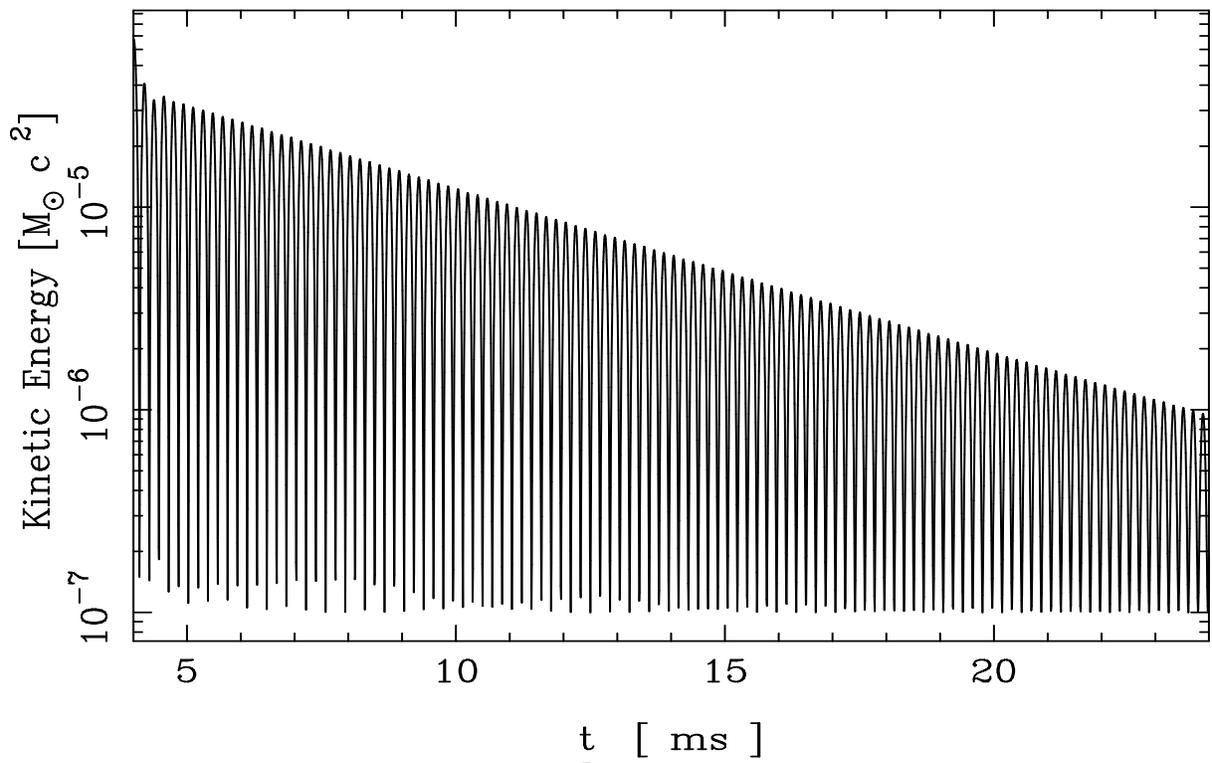}}
\caption[]{\label{f:varecin}
Evolution of star's kinetic energy defined as $\int_{r=0}^{r=R}4\pi
\Gamma(\Gamma -1)\tilde{e}r^2\rm{d}r$, where $\tilde{e}$ is the total
energy density of fluid, for unstable initial equilibrium
solution number 4 (Cf. Table~\ref{t:model}) .}
\end{figure}

This last point is studied more in details. There is no rigorous
definition of the kinetic energy of a star in general relativity,
however if one considers the kinetic energy of a particle of mass $m$:
$$
E^0_{kin} = (\Gamma - 1) m c^2,
$$
$\Gamma$ being its Lorentz factor, then a good choice of $E_{kin}$ for
a star may be given by:
\be
E_{kin}=\int_{r=0}^{r=R}4\pi \Gamma(\Gamma -1)\tilde{e}r^2\rm{d}r,
\label{e:defekin}
\ee
with $\Gamma \tilde{e}$\footnote{the additional Lorentz factor is due
to the relativistic contraction of space} replacing $mc^2$. $E_{kin}$
is thus computed the same way as in \cite{GHG95}. 
On Fig.~\ref{f:varecin} the evolution of this kinetic energy, giving
the good Newtonian limit, 
  is
plotted and shows well the damped behavior (exponential decay). Such
neutron stars are really 
damped oscillators, except for stars with masses close to the
critical one (described in Fig.~\ref{f:sseos}) which exhibit no
oscillations, but only exponential relaxation toward equilibrium
state. In any case, the maximal velocity reached by star's surface is
of the order of $10^{-2} c$ showing clearly that the appearing strong
scalar field has an important influence on the structure and the
hydrodynamics of the star.

\section{Gravitational emission during the transition}
\label{s:gw}

\subsection{Monopolar component}\label{ss:mono}

\begin{figure}
\centerline{\epsfig{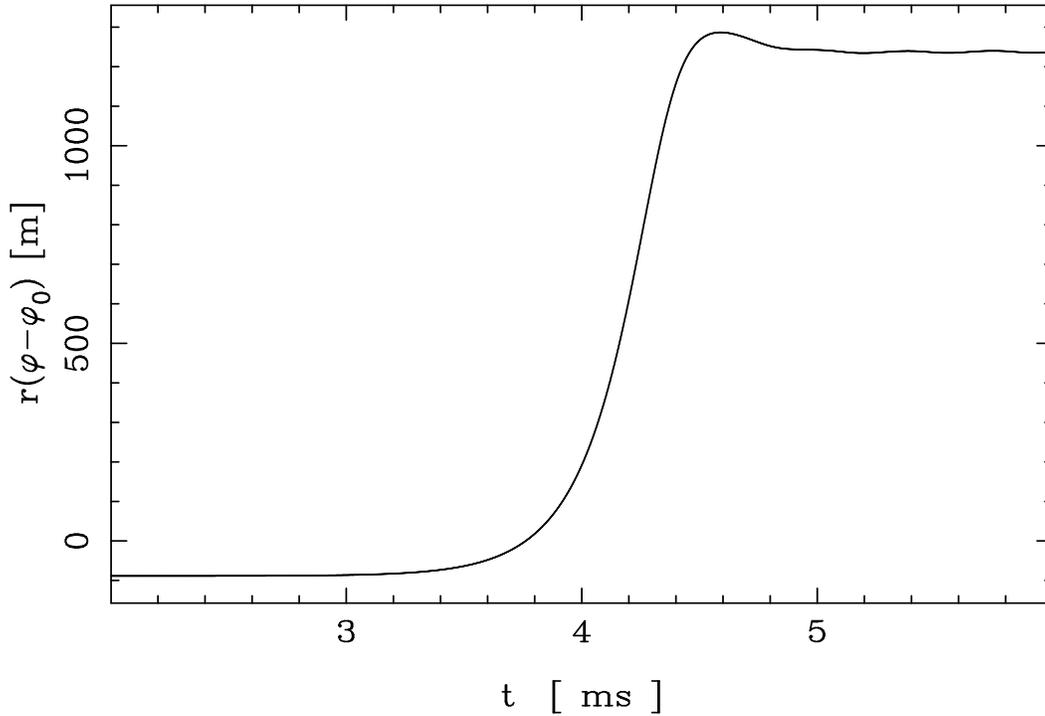}}
\caption[]{\label{f:varfio1}
Scalar waveform resulting from the evolution of unstable initial equilibrium
solution number 4 (Cf. Table~\ref{t:model}), corresponding to the
raise of the scalar field.}
\end{figure}

\begin{figure}
\centerline{\epsfig{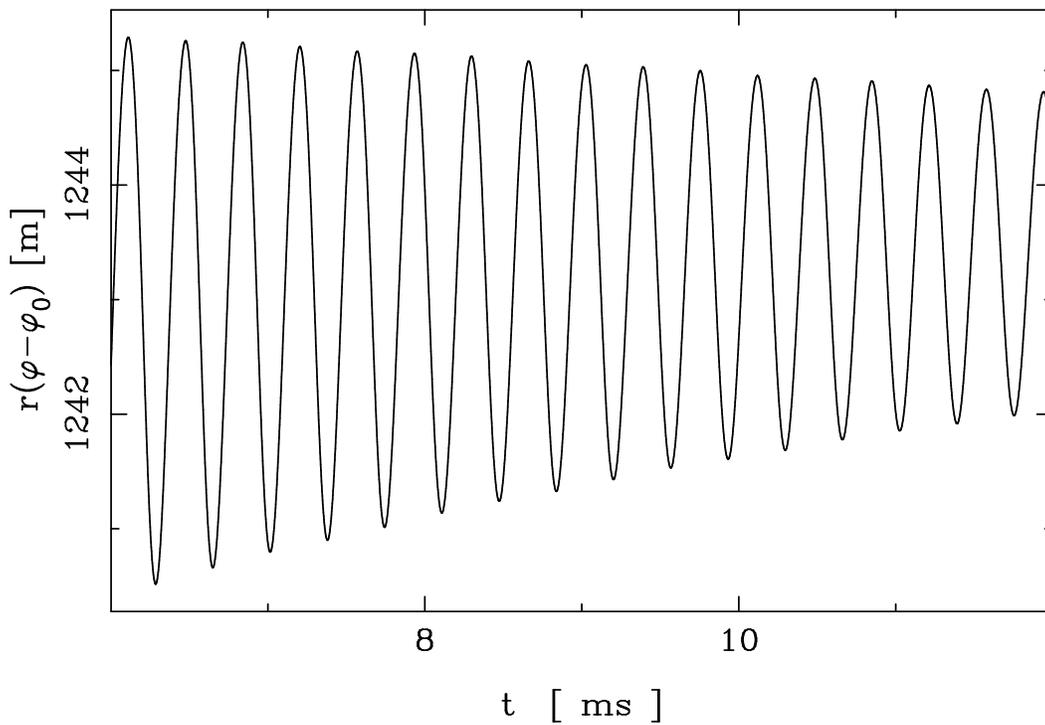}}
\caption[]{\label{f:varfio2}
Scalar waveform resulting from the evolution of unstable initial equilibrium
solution number 4 (Cf. Table~\ref{t:model}), corresponding to the
damping of star's oscillations.}
\end{figure}

\begin{figure}
\centerline{\epsfig{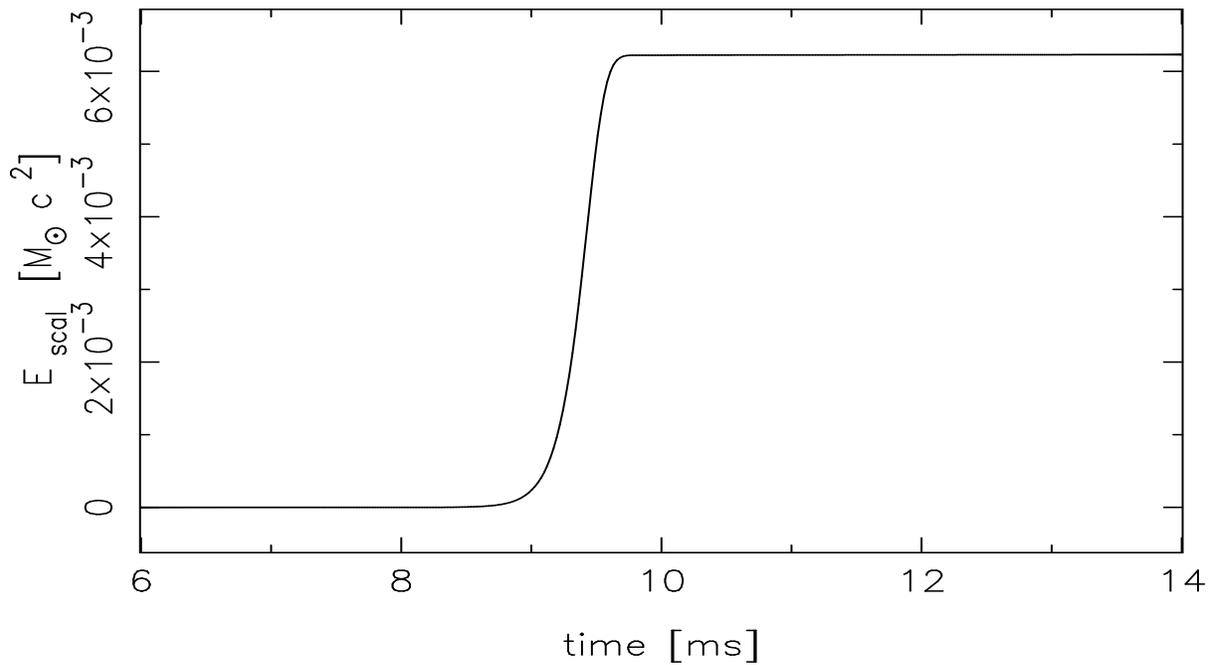}}
\caption[]{\label{f:vareray}
Evolution of star's radiated energy defined in eq.(~\ref{e:eray}), for
unstable initial equilibrium solution number 4 (Cf. Table~\ref{t:model}) .}
\end{figure}

Variation of the scalar field during the transition results in an
emission of monopolar gravitational waves. Far from the star, the
scalar field writes:
\be
\varphi (r,t) = \varphi_0 + {1\ov r}F\l(t-{r\ov c}\r) +
O\l({1\ov r^2}\r), \label{e:filoin}
\ee
and the interesting ``scalar wave'' which can be detected by an
gravitational wave detector, at a distance $d$ from the source, is then
related to $F$ through: 
\be
h(t) = {2 \ov d} \alpha_0 F(t)
\label{e:ampli}
\ee
(see eq.(5.6) of \cite{DEF92}, with $a(\varphi_0)=1$).
The function $F(t)$ (waveform) is plotted in Figs.~\ref{f:varfio1} and
\ref{f:varfio2}, for the rapid variation and the damped oscillations
respectively. The first step has a characteristic frequency of $\sim
200$~Hz, whereas the second one, lower in amplitude, has $\sim
3$~kHz. These results depend essentially on the star's mass. From these
waveforms, one can also deduce the radiated scalar energy, defined as:
\be
E_{scal} = {c^3\ov G_*} \int_0^{+\infty} \l({dF\ov dt}\r)^2 dt, \label{e:eray}
\ee
which is plotted in Fig.~\ref{f:vareray} as a function of time. The
function $F(t)$ is estimated at $r=300$ km, see \cite{NOV97}. The
radiated amount corresponds (within errors) to the difference between
static cases 4 and 5 of table~\ref{t:model}. On Fig.~\ref{f:vareray},
the star still radiates 
some energy after $t=10$ ms: it is its kinetic energy of
Fig.~\ref{f:varecin}. If one does not consider this oscillations (which are
small compared to the first raise), the wave represents a transition
which is the inverse of that of neutron star collapse to a black hole,
when the star is strongly scalarized. The scalar field goes from the
strong-field value to the asymptotic one, $\varphi_0$, whereas in
Fig.~\ref{f:varfio1} it goes from (almost) $\varphi_0$ to its strong field
value. Therefore, the amplitude of the wave resulting from a transition
of a neutron star to a strong-scalar-field state is very similar to
that of a wave coming from a neutron star collapse to a black
hole. The discussion on this amplitude of the wave made in 
\cite{NOV97} can be applied here to say that the monopolar radiation
from such events is unlikely to be observed, with constraints on the
$\alpha_0 , \beta_0$ parameters taken from solar system experiments
\cite{DEF96}, if the sources are 
situated at a larger distance than a few $100$ kpc. This excludes the
Virgo cluster galaxies and thus reduces the number of possible
sources. 

\subsection{Quadrupolar component}\label{ss:quad}

From eq.~\ref{e:ampli} one sees that the amplitude of the wave
interacting with the detector is directly proportional to
$\alpha_0$. Since cosmological arguments (\cite{DN93} and \cite{DN93b})
indicate that the parameter $\alpha_0$ should have been driven to $0$
by cosmological evolution, monopolar waves described in the previous
section can be
``invisible'' to gravitational detectors. However, if one considers
a slowly-rotating neutron star, which is very close to spherical
symmetry, the oscillations of its surface (Figs.~\ref{f:varray}
and \ref{f:varv}) will induce a modification of its quadrupolar
momentum, the star will emit (usual) quadrupolar gravitational
radiation, whose detection is not sensitive to $\alpha_0$. This qualitative
scenario should also hold for rapidly-rotating neutron stars, but the
aim of this section is to make an order-of-magnitude crude estimate of
the emitted quadrupolar wave amplitude. The detection of such
quadrupolar waves, with no detection of monopolar ones, would give
solid constraints on the theory, indicating the possibility for
spontaneous scalarization and constraining $\alpha_0$. 

Therefore, the procedure is very simplified: at each time-step of the
dynamical integration the star is supposed to be at equilibrium. As a
slowly-rotating star, it can be described by the 
perturbation equations of Hartle \cite{HAR67} and Hartle and Thorne
\cite{HAT68}, which are second order accurate in the angular
velocity. Their work starts from a spherically symmetric static star,
described in the same gauge as in this work, which is perturbed by the
rotation. The resulting metric involves several additional functions,
proportional to angular velocity or to its square: 
\begin{eqnarray}
ds^2 &=& - N^2[1+2(h_0+h_2 P_2)]dt^2 + \l[1+{2\ov r}A^2(m_0+m_2
P_2)\r] dr^2 \nonumber\\
&+&  r^2[1+2(v_2-h_2)P_2][d\theta^2 + \sin^2 \theta(d\phi -
\omega dt)^2] + O(\Omega^3), \label{e:mrot}
\end{eqnarray}
with $\Omega$ being star's angular velocity, $\omega(r)$ the ``angular
velocity of the local inertial frame'', $A$ and $N$ are defined in
eq.~(\ref{e:metric}), $P_2=P_2(\cos\theta)$ is the second Legendre
polynomial and $h_0, h_2, m_0, m_2, v_2$ are perturbation functions
proportional to $\Omega^2$. From the asymptotic behavior of these last
functions, one can deduce the quadrupolar momentum. Since all the
quantities are computed at each time-step, from the matter
distribution and fields, one also knows the quadrupolar momentum $Q$
of the star at each time-step. One then uses the usual quadrupole
formula to get the amplitude of the wave at a distance $d$:
\be
h_{TT}(t) = {2G_*\ov c^4 d} {d^2Q(t) \ov dt^2}.
\ee
The value obtained for $Q$ from the simulation described in
sec.~\ref{s:ssdyn} is:
$$
{d^2Q\ov dt^2} \sim 10^{32}\times f^2 \, {\rm kg.m}^2{\rm .s}^{-2},
$$
with $f$ being the rotation frequency of the star, expressed in
Hz. This yields 
\be
h_{TT} \sim 10^{-12}\times {f^2\ov d}.
\ee
So, even if the highest value for the rotation frequency is considered
($2$~{kHz}, see \cite{SBG94}), if this wave is to be detected by the
interferometers currently under construction (whose detection level is
about $10^{-23}$ at these frequencies), the source has to be situated
closer than $10$~{pc}, that is very close in our Galaxy.

\section{Conclusions} \label{s:conc}

Spontaneous scalarization effects appear in a restricted part of the
($\beta_0$,$M_B$) parameter space, depending on the equation of state
used for neutron star matter. On the contrary, the maximal $\beta_0$
parameter which allows for spontaneous scalarization is
quasi-independent of the equation of state, its value being
$-4.34$. When these types of equilibrium solutions exist, they are
energetically more favorable than the weak-field ones, which are then
unstable equilibria. A star which is in the weak-field regime (when its mass
is to low to get ``scalarized'') can increase the mass by accretion
and thus become unstable. Then it undergoes evolution to the
strong-field state  radiating the difference of energy as monopolar
gravitational wave; it also behaves like a damped oscillator, since a
small fraction of the energy ($\sim 10^{-5}M_\odot c^2$) is put as
kinetic oscillating energy into 
the star. The transitions studied in this
work are not likely to be observed through their gravitational
radiation, excepted if they are located close to us (which severely
decreases the number of possible sources). However, the kinetic energy put
into the star has an important influence on the
hydrodynamics and its effect must not be neglected when considering
the supernovae collapse toward a neutron star: when the proto-neutron
star gets its final compactness the scalar field can develop and
influence the ejection of the envelop. The future study of supernova
collapse and bounce, within the framework of tensor-scalar theory
could provide us with monopolar gravitational signals. These latter
can give constraints on the tensor-scalar theory space parameter, even
if they are not detected, since in the case of a supernova,
electromagnetic or neutrino signals are detected.

\newpage

\newpage

\begin{table}
\caption{\label{t:model}
Computed characteristics of different neutron stars of $1.5 M_\odot$ showing
spontaneous scalarization effects (solutions 2,3,5 and 6) or without
(solutions 1 and 4). The equation of state (EOS1) is described in
sec.~\ref{ss:coord}, $\alpha_0$ and $\beta_0$ are the
coupling function parameters (\ref{e:cfunc}), $R_{star}$ denotes
star's radius, $\tilde{n}_B(r=0)$ is the central baryon density (in
units of nuclear density, $1\ n_{nuc}=10^{44}\ \text{m}^{-3}$), $M_G$
is the $g^*_{\mu\nu}$-frame ADM mass, $M_B$ the baryonic one and
$\omega$ the scalar charge. The code used for these results is a static one.}
\begin{tabular}{c|c|c|c|c|c|c|c}
 Solution & $\alpha_0$ & $\beta_0$ & $R_{star}$ &
 $\tilde{n}_B(r=0)$ & $M_G$ & $M_B$ & $\omega$ \\
&&& [km] & [$n_{nuc}$] & [$\ M_\odot$] & [$\ M_\odot$] & [$\
 M_\odot$] \\
\hline
 1 & $0$ & $-6$ & $13.2$  &
 $3.9643$ & $1.37803$ & $1.50009$ & $0$ \\

 2 & $0$ & $-6$ & $13$ &
 $5.3212$ & $1.37322$ & $1.50008$ & $0.781$ \\

 3 & $0$ & $-6$ & $13$ &
 $5.3212$ & $1.37322$ & $1.50008$ & $-0.781$ \\

 4 & $ 10^{-2}$ & $-6$ & $13.2$  & $3.9742$ &
 $1.37807$ & $1.50007$ & $-5.91\times 10^{-2}$ \\

 5 & $ 10^{-2}$ & $-6$ & $13$ & $5.3674$
& $1.3719$ & $1.50008$ & $0.803$ \\

 6 & $ 10^{-2}$ & $-6$ & $13$ & $5.2669$
& $1.37452$ & $1.50008$ & $-0.757$ \\
\end{tabular}
\end{table}

\end{document}